\documentclass[%
 preprint,
 superscriptaddress,
 amsmath,amssymb,
 aps,physrev,
 floatfix,
]{revtex4-2}

\usepackage[T1]{fontenc}
\usepackage[utf8]{inputenc}
\usepackage{graphicx}
\usepackage{dcolumn}
\usepackage{bm}

\newcommand{\FeRhSi}{Fe$_{0.5}$Rh$_{0.5}$Si}
\newcommand{\FeCoSi}{Fe$_{1-x}$Co$_x$Si}

\begin{document}


\title{SANS and magnetometry study of the magnetic phase diagram of the B20 helimagnet \FeRhSi}

\author{E. V. Altynbaev}
\affiliation{Saint Petersburg State University, St. Petersburg, Russia}
\affiliation{Vereshchagin Institute for High Pressure Physics, Russian Academy of Sciences, 108840 Troitsk, Moscow, Russia}

\author{A. V. Guseva}
\affiliation{Saint Petersburg State University, St. Petersburg, Russia}

\author{D. O. Skanchenko}
\affiliation{Mozhaisky Military Space Academy, St. Petersburg, Russia}
\affiliation{Vereshchagin Institute for High Pressure Physics, Russian Academy of Sciences, 108840 Troitsk, Moscow, Russia}

\author{V. N. Krasnorussky}
\affiliation{Vereshchagin Institute for High Pressure Physics, Russian Academy of Sciences, 108840 Troitsk, Moscow, Russia}

\author{A. V. Bokov}
\affiliation{Vereshchagin Institute for High Pressure Physics, Russian Academy of Sciences, 108840 Troitsk, Moscow, Russia}

\author{D. A. Salamatin}
\affiliation{Vereshchagin Institute for High Pressure Physics, Russian Academy of Sciences, 108840 Troitsk, Moscow, Russia}

\author{V. A. Sidorov}
\affiliation{Vereshchagin Institute for High Pressure Physics, Russian Academy of Sciences, 108840 Troitsk, Moscow, Russia}

\author{A. V. Tsvyashchenko}
\affiliation{Vereshchagin Institute for High Pressure Physics, Russian Academy of Sciences, 108840 Troitsk, Moscow, Russia}

\date{\today}

\begin{abstract}
\FeRhSi{} was recently identified as a 4d-substituted B20 helimagnet. Here we report a combined low-field magnetometry and small-angle neutron scattering (SANS) study of a newly synthesized polycrystalline sample. Regularized magnetization isotherms $M(H)$ were used to construct a low-field $H$-$T$ phase diagram; the magnetometric ordering scale $T_C^{\mathrm{mag},H_{c2}}=71\pm2$ K is defined as the temperature at which the characteristic fields are expected to extrapolate to zero. The characteristic fields $H_{c1}$, $H_{c1m}$, and $H_{c2}$ respectively mark the onset of field-induced helical-domain reorientation, the completion of the main reorientation process, and the crossover toward the field-polarized state. Within the $H_{c1}$-$H_{c2}$ interval, the differential susceptibility shows additional anomalies that define a continuous candidate A-phase region spanning approximately 56--68 K from magnetometry. SANS confirms long-period helimagnetic order with $k_s=0.00793$ \AA$^{-1}$, $\lambda_h\approx79$ nm, and a SANS-derived magnetic ordering scale $T_C^{\mathrm{SANS}}\approx70\pm5$ K. Field-dependent SANS shows a perpendicular-to-field SANS-intensity maximum near 60 K at $\mu_0 H=0.025\pm0.005$ T, which we use as a SANS-derived marker for the candidate A-phase region. The SANS data support the candidate A-phase region; however, an unambiguous determination of skyrmion-lattice structural parameters requires further experimental studies.
\end{abstract}

\maketitle

\section{Introduction}

Cubic B20 magnets provide a particularly clean setting in which crystallographic chirality and relativistic spin-orbit coupling generate long-wavelength magnetic order. In the noncentrosymmetric $P2_13$ structure, ferromagnetic exchange favors locally parallel moments, the Dzyaloshinskii-Moriya interaction (DMI) fixes a finite helical wave vector, and much weaker anisotropies select preferred propagation directions \cite{bak1980,dzyaloshinsky1958,moriya1960}. This hierarchy, formulated for MnSi and FeGe by Bak and Jensen \cite{bak1980}, remains the standard framework for interpreting helimagnetic and conical states in B20 silicides and germanides.

The B20 family is also central to the modern literature on skyrmion-lattice and A-phase physics. In MnSi, SANS established the skyrmion lattice in the A phase \cite{muhlbauer2009}, while related work identified \FeCoSi{} as a chemically substituted B20 skyrmion host \cite{munzer2010}. These studies showed that the A phase is a narrow field-temperature region near the ordering temperature, but they also demonstrated that the experimental signature depends on sample form and SANS geometry. In single crystals with the field parallel to the incident neutron beam, a skyrmion lattice may appear as a sixfold diffraction pattern. In transverse geometry, where the field is perpendicular to the incident beam, additional magnetic Bragg spots transverse to the field direction can also indicate an A-phase or multidomain skyrmion-lattice state, beyond the two conical spots aligned with the field. In polycrystalline samples, the corresponding signal may appear as a field-induced ring or redistribution of intensity rather than a clean sixfold pattern \cite{muhlbauer2009,munzer2010,muhlbauer2019}. Thus, a magnetometric candidate A-phase anomaly in \FeRhSi{} should be regarded as a candidate region until it is directly verified by SANS in the appropriate geometry, Lorentz transmission electron microscopy (LTEM), magnetic force microscopy (MFM), or topological Hall measurements.

\FeCoSi{} is the closest established reference system for \FeRhSi. Neutron and magnetization studies established the composition dependence of the transition temperature, helix wave vector, and principal magnetic interactions \cite{grigoriev2007brief}, while polarized SANS in field resolved the evolution from helical domains through the conical state toward the field-polarized regime \cite{grigoriev2007field}. Later work emphasized that the A-phase pocket, metastable textures, and apparent phase boundaries can depend strongly on disorder, magnetic history, and thermal protocol \cite{munzer2010,bauer2016}. This background is directly relevant for a Rh-substituted B20 silicide, where chemical disorder and stronger 4d spin-orbit coupling may both influence the balance of exchange, DMI, and anisotropy.

\FeRhSi{} was recently synthesized and reported as a new chiral B20 helimagnet \cite{tsvyashchenko2026}. That work established the B20 structural platform and reported bulk magnetic, transport, heat-capacity, NMR, and electronic-structure signatures consistent with helimagnetism, with a transition scale near $T_C=65.6$ K and a spontaneous moment of order $0.48\,\mu_B$ per Fe \cite{tsvyashchenko2026}. Rh substitution is also motivated by recent work on Rh-doped MnSi, where NMR and electronic-structure analysis revealed disorder-induced coexistence of itinerant low-spin and localized high-spin Mn states \cite{krasnorussky2024}. These results make \FeRhSi{} a useful platform for testing how 4d substitution modifies B20 helimagnetism. However, before the present study, the long-period magnetic modulation in \FeRhSi{} had not been directly established by SANS.

Small-angle neutron scattering is the appropriate bulk probe for this problem because the expected helix periods in B20 magnets lie on mesoscopic length scales. Magnetic SANS directly measures the ordering wave vector, distinguishes ring-like multidomain helical scattering from field-selected textures, and can track the redistribution or suppression of magnetic Bragg intensity through $H_{c1}$, $H_{c1m}$, and $H_{c2}$ \cite{muhlbauer2019,grigoriev2007field}. In the present work, we combine low-field magnetometry and SANS on \FeRhSi. Magnetometry is used to construct the low-field $H$-$T$ phase diagram and to locate a continuous candidate A-phase region, while SANS confirms long-period helimagnetic order and provides an independent structural check of the field-induced intensity redistribution.

The SANS measurements give $k_s=0.00793$ \AA$^{-1}$, corresponding to $\lambda_h\approx79$ nm, and a SANS-derived magnetic ordering scale $T_C^{\mathrm{SANS}}\approx70$ K. Field-dependent SANS shows redistribution and suppression of magnetic Bragg intensity across the low-field phase boundaries, including a perpendicular-to-field SANS-intensity maximum near the magnetometrically defined candidate A-phase region. We interpret this observation as supporting evidence for the candidate A-phase region, but an unambiguous determination of a skyrmion-lattice state and its structural parameters will require additional experiments in an appropriate field geometry or complementary probes.

\section{Magnetometry experiments and analysis}

The synthesis and primary structural, transport, thermodynamic, NMR, and DFT characterization of \FeRhSi{} were reported previously \cite{tsvyashchenko2026}. The sample investigated here was obtained in a separate high-pressure, high-temperature synthesis that reproduced the preparation procedure described in Ref.~\cite{tsvyashchenko2026}. A powder x-ray diffraction check of a representative fragment from the present synthesis was consistent with the cubic B20 phase and gave a lattice parameter close to $a=4.61$ \AA, in agreement with the range reported in Ref.~\cite{tsvyashchenko2026}; the detailed structural characterization and Rietveld analysis are therefore not repeated here. The material was polycrystalline, and no crystallographic orientation was assigned in the present analysis. Magnetometry and SANS were performed on polycrystalline specimens from the same preparation series, with sample shape and mass chosen for the requirements of each technique.

Magnetization measurements and measurement protocols followed Ref.~\cite{tsvyashchenko2026}, using a Quantum Design PPMS-9 magnetometer. Isothermal $M(H)$ curves were measured after zero-field cooling to the target temperature, following the demagnetization and field-sweep procedures described in Ref.~\cite{tsvyashchenko2026}. The low-field analysis in this work used the first increasing-field branch of these isotherms.

For each temperature, a regularized smooth representation of $M(H)$ was constructed in the low-field window, and the differential susceptibility $\chi(H)=dM/dH$ and curvature $d^2M/dH^2$ were evaluated from that model. $H_{c1}$ is defined as the first robust maximum of $d^2M/dH^2$, marking the onset of low-field helical-domain reorientation. $H_{c1m}$ is the following minimum of $d^2M/dH^2$ and is associated with the completion of the main domain-reorientation process, after which no further growth of the field-aligned helical/conical contribution is observed and the magnetic signal decreases toward $H_{c2}$. $H_{c2}$ is taken on the high-field recovery branch after the last high-field minimum in $d^2M/dH^2$. Where useful, the $H_{c2}$ corridor denotes the interval between that minimum and the extrapolated high-field recovery point at which the magnetic response is expected to approach the field-polarized saturation trend, with the first- and second-derivative trends tending to zero within the adopted regularized model.

The candidate A-phase anomaly was searched only after the main phase boundaries were fixed. The search was carried out within the broader $H_{c1}<H<H_{c2}$ interval, rather than by assuming a sharp boundary at $H_{c1m}$. In practice, the most relevant features occur in the relatively high-field part of this interval, where the comparison with SANS indicates that the main domain-reorientation process is largely completed. The use of susceptibility features is guided by the established A-phase phenomenology in MnSi: in the A phase the differential susceptibility is reduced relative to the neighboring conical response, while the adjacent transition region is associated with a recovery of susceptibility and a peak-like feature in $dM/dH$; neutron diffractive imaging has further shown that such bulk features can reflect a coexistence/transition region near the skyrmion-lattice boundary \cite{reimann2018}. We therefore use as an empirical marker a positive lobe of the regularized $d^2M/dH^2$, corresponding to an interval in which $\chi(H)$ increases between $H_{A1}$ and $H_{A2}$ after a local depression of the differential susceptibility. The continuous region bounded by these features over approximately 56--68 K is treated as a magnetometric indicator of a possible A-phase region, not as a precise thermodynamic phase boundary.

\section{Magnetometry: low-field phase boundaries and candidate A-phase region}

\begin{figure}[htbp]
\centering
\includegraphics[width=\textwidth]{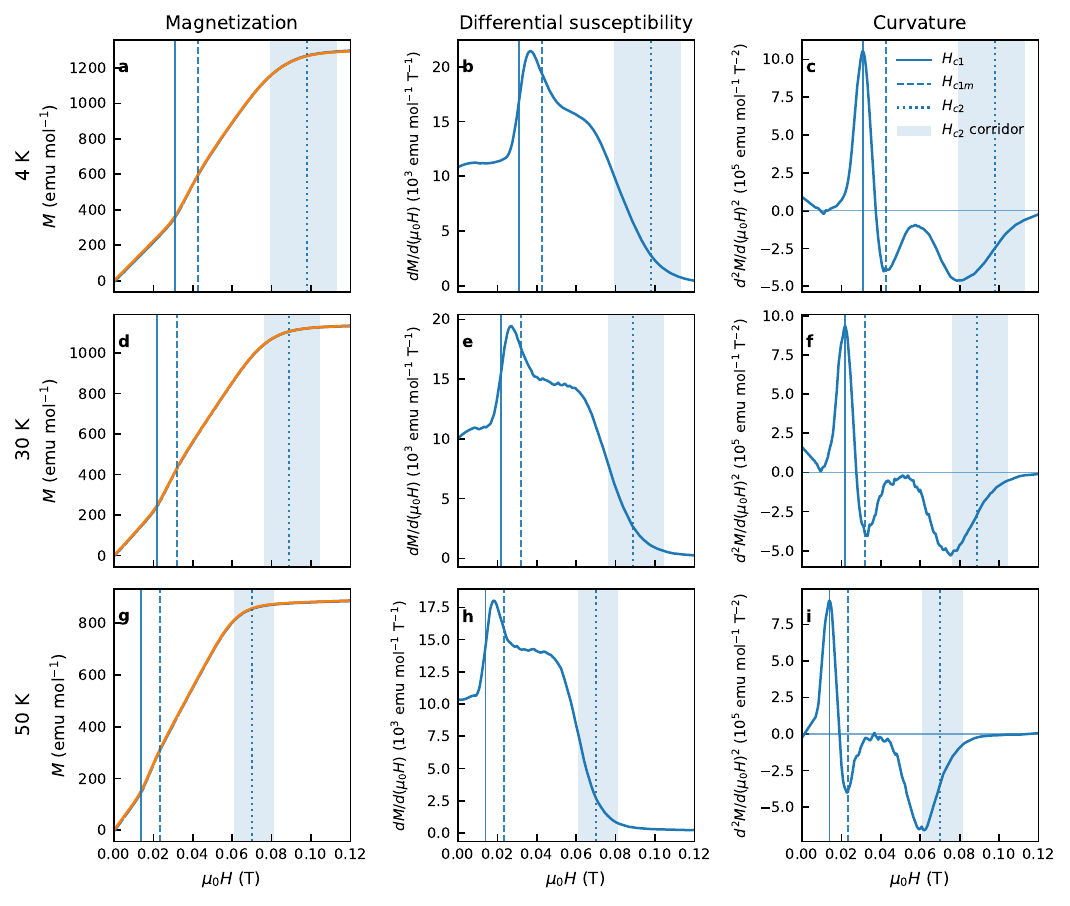}
\caption{Representative magnetometry plots from $M(H)$ isotherms at 4, 30, and 50 K. Each row shows $M(H)$, $\chi(H)=dM/dH$, and regularized $d^2M/dH^2$ for one temperature. Vertical markers denote $H_{c1}$, $H_{c1m}$, and $H_{c2}$; the shaded interval denotes the $H_{c2}$ corridor used in the magnetometric phase-boundary extraction.}
\label{fig:magnetometry-matrix}
\end{figure}

Representative magnetometry diagnostics are shown in Fig.~\ref{fig:magnetometry-matrix}. Although the $M(H)$ curves are smooth in the low-field range, the regularized derivatives reveal a reproducible sequence of extrema that defines the characteristic fields used in the phase diagram. The first robust positive-curvature maximum is assigned to $H_{c1}$, the subsequent minimum to $H_{c1m}$, and the high-field recovery feature to $H_{c2}$. In this convention, the three lines describe successive stages of the field-driven evolution from multidomain helical order to the field-polarized regime.

\begin{figure}[htbp]
\centering
\includegraphics[width=\textwidth]{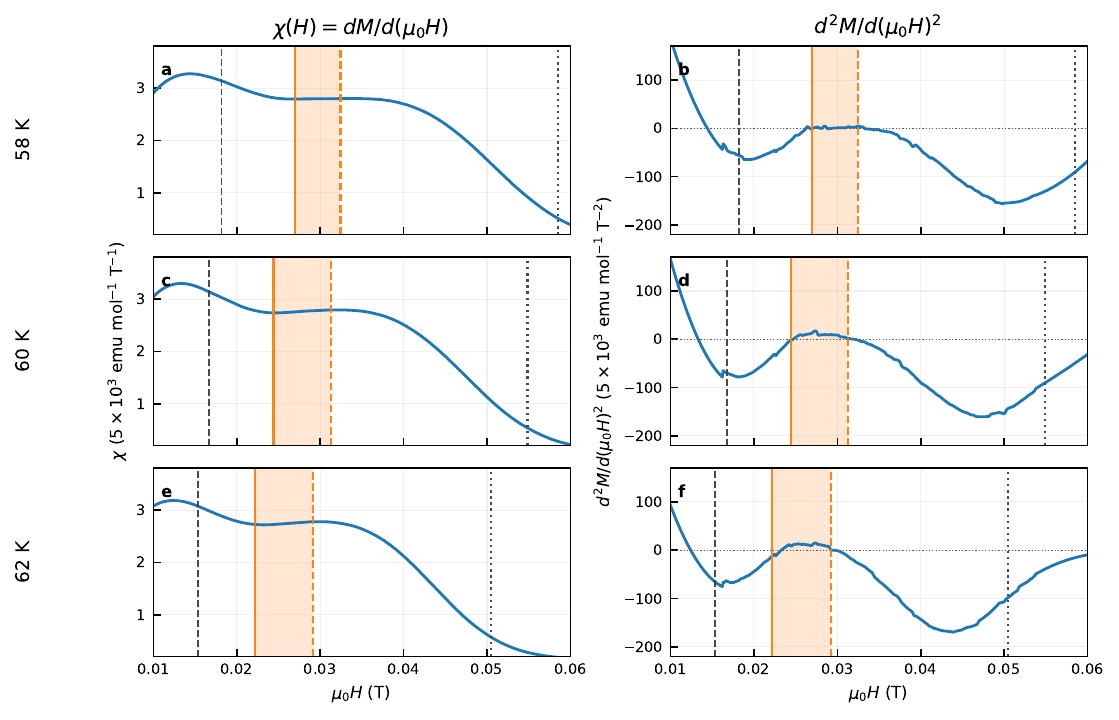}
\caption{Candidate A-phase diagnostics at 58, 60, and 62 K. Left: $\chi(H)=dM/d(\mu_0H)$; right: $d^2M/d(\mu_0H)^2$. Shading marks $H_{A1}$-$H_{A2}$; dashed black, orange solid/dashed, and dotted black lines mark $H_{c1m}$, $H_{A1}/H_{A2}$, and $H_{c2}$.}
\label{fig:aphase-diagnostics}
\end{figure}

After the main phase boundaries were fixed, the $H_{c1}$-$H_{c2}$ interval was examined for additional structure in $\chi(H)$. Figure~\ref{fig:aphase-diagnostics} illustrates the way in which the candidate A-phase boundaries are defined: $\chi(H)$ develops a shallow depression followed by a local recovery, producing a positive curvature lobe in $d^2M/dH^2$ between $H_{A1}$ and $H_{A2}$. The detailed comparison with the field-dependent SANS data shows that these features occur in the relatively high-field part of the low-field phase diagram, where the perpendicular-to-field SANS intensity also has its strongest response. Taken together over neighboring temperatures, these paired features define a continuous candidate A-phase region over approximately 56--68 K.

The anomaly occupies a restricted field interval inside $H_{c1}$-$H_{c2}$. We therefore refer to it as a candidate A-phase region. This wording is deliberately conservative: the magnetometric feature is consistent with the A-phase phenomenology of other B20 helimagnets and is supported by the perpendicular-to-field SANS-intensity maximum discussed below, but it is not by itself a direct identification of a skyrmion lattice. Following susceptibility-based analyses of the A phase and adjacent transition/coexistence regions in MnSi \cite{reimann2018}, we use the positive-curvature interval as a restrictive marker of where such a state may occur, rather than as an unambiguous boundary of its full stability range. The low-field side of this region may be especially sensitive to crystallographic orientation, demagnetizing fields, and sample geometry; for the present polycrystalline sample the shaded region should therefore be read as a conservative experimental constraint. An unambiguous skyrmion-lattice assignment would require additional reciprocal-space evidence, such as a clear multidomain or sixfold SANS pattern in the appropriate field geometry, or complementary real-space/topological Hall measurements.

\begin{figure}[htbp]
\centering
\includegraphics[width=\textwidth]{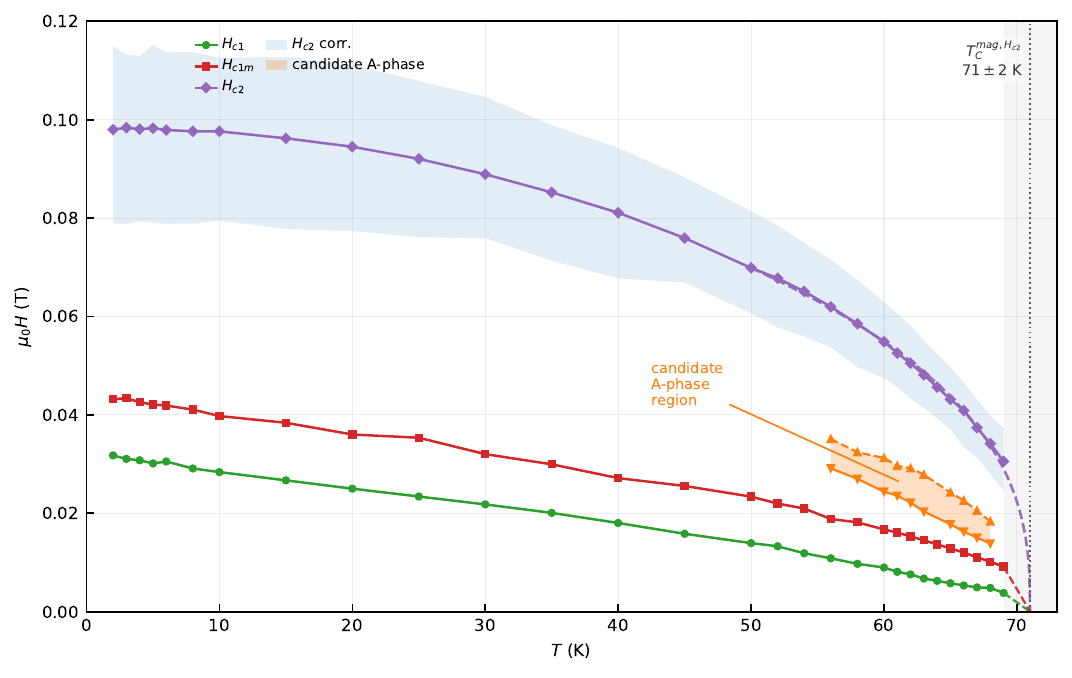}
\caption{Magnetometric $H$-$T$ phase diagram constructed from the low-field derivative criteria. Symbols show $H_{c1}$, $H_{c1m}$, and $H_{c2}$; the blue band is the $H_{c2}$ corridor between the high-field curvature minimum and the extrapolated high-field recovery point. The orange band marks the candidate A-phase region bounded by $H_{A1}$ and $H_{A2}$. The dashed $H_{c2}$ curve is a power-law fit constrained to vanish at $T_C^{\mathrm{mag},H_{c2}}=71$ K; dashed $H_{c1}$ and $H_{c1m}$ segments are visual guides to zero field.}
\label{fig:magnetometric-phase-diagram}
\end{figure}

The resulting magnetometric phase diagram is shown in Fig.~\ref{fig:magnetometric-phase-diagram}. $H_{c1}$, $H_{c1m}$, and $H_{c2}$ decrease with increasing temperature, and the $H_{c2}$ corridor indicates the spread between the two high-field extraction criteria. In the present data set the $H_{c2}$ line approaches zero field at $T_C^{\mathrm{mag},H_{c2}}=71\pm2$ K. This value is used below as a characteristic low-field magnetometric scale, not as a model-independent thermodynamic Curie temperature, and is distinct from the published bulk $T_C=65.6$ K for an independently synthesized sample.

\section{SANS experiments and evidence for helimagnetism}

Small-angle neutron scattering experiments were carried out on the BL01 small-angle neutron scattering instrument at the Chinese Spallation Neutron Source \cite{ke2018} using a polycrystalline sample. Time-of-flight wavelengths from 1.0 to 9.15 \AA{} were used, with a sample-to-detector distance of 5 m; these settings were chosen to cover the low-$Q$ region containing the expected helical wave vector near 0.008 \AA$^{-1}$. The sample was mounted in a horizontal-field magnet in transverse geometry, with the incident neutron beam perpendicular to the applied field. Corrected two-dimensional detector maps were reduced to radial or sector-averaged one-dimensional profiles after detector-efficiency correction and background subtraction.

\begin{figure}[htbp]
\centering
\includegraphics[width=\textwidth]{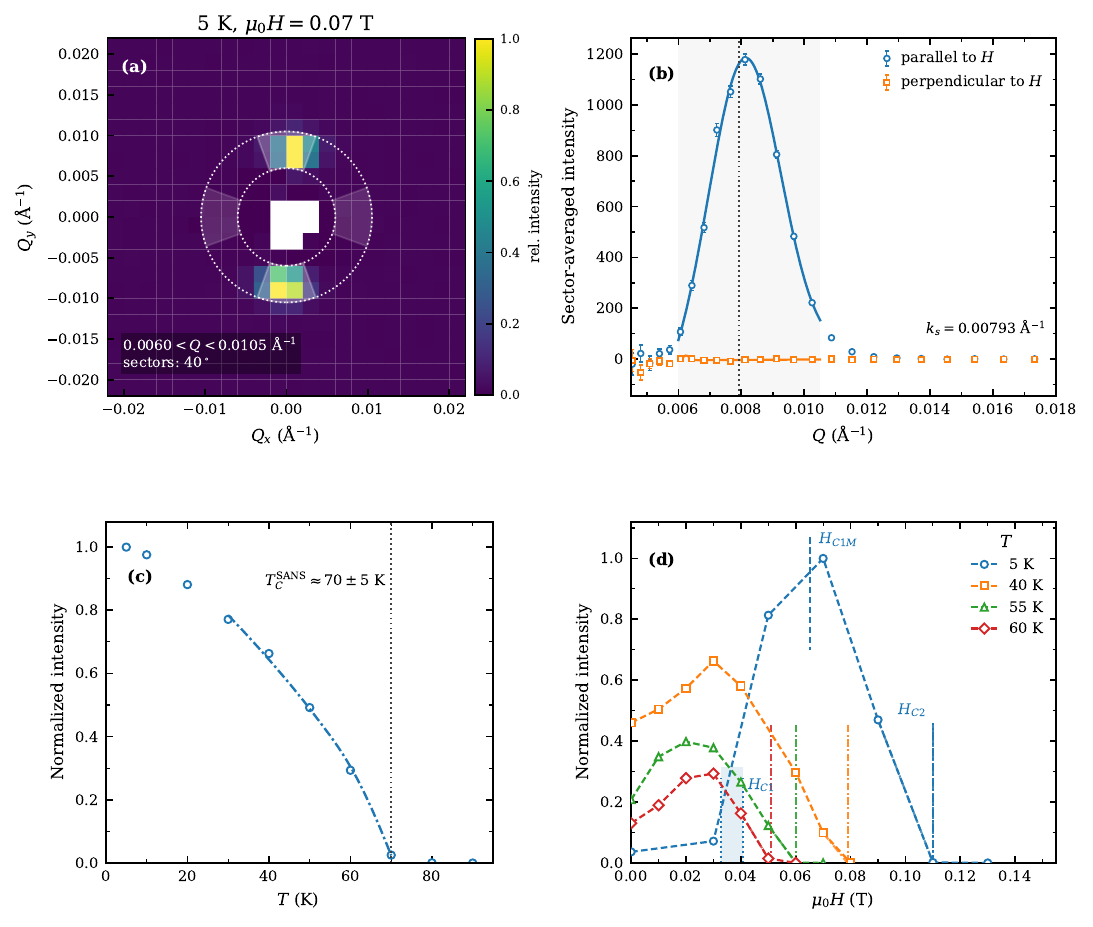}
\caption{SANS evidence for helimagnetic order and field evolution. (a) Corrected 2D SANS map at 5 K and $\mu_0H=0.07$ T; shaded sectors indicate the parallel-to-$H$ and perpendicular-to-$H$ azimuthal-integration windows, corresponding respectively to the vertical and horizontal directions on the detector, in the annulus $0.0060<Q<0.0105$ \AA$^{-1}$. (b) Sector-averaged $I(Q)$ profiles for the same geometry; symbols are data, curves are Gaussian fits on a linear background, and the dotted vertical line marks $k_s=0.00793$ \AA$^{-1}$. (c) Normalized integrated SANS intensity from the zero-field temperature scan; the dash-dotted curve and dotted vertical line indicate the estimate $T_C^{\mathrm{SANS}}\approx70\pm5$ K. (d) Normalized field-dependent SANS intensity at selected temperatures; dashed lines connect the data, dash-dotted markers indicate $H_{c2}$ estimates obtained from high-field tail fits, and the 5 K trace illustrates $H_{c1}$, $H_{c1m}$, and $H_{c2}$.}
\label{fig:sans}
\end{figure}

Small-angle neutron scattering results are summarized in Fig.~\ref{fig:sans}. The corrected 2D map at 5 K and $\mu_0H=0.07$ T in Fig.~\ref{fig:sans}(a) shows anisotropic magnetic scattering in transverse geometry, and the overlaid sectors define the parallel-to-$H$ and perpendicular-to-$H$ integration windows. In the detector coordinates these correspond to the vertical and horizontal directions, respectively. The corresponding sector-averaged $I(Q)$ profiles in Fig.~\ref{fig:sans}(b) show a magnetic peak centered at $k_s=0.00793\pm0.00006$ \AA$^{-1}$, giving a helical period $\lambda_h\approx79$ nm. This is the central structural result: SANS directly confirms the long-period helimagnetic modulation in \FeRhSi.

The temperature dependence of the SANS signal is shown in Fig.~\ref{fig:sans}(c). The normalized integrated intensity decreases steadily on warming and approaches the background near 70 K, defining a SANS-derived ordering scale $T_C^{\mathrm{SANS}}\approx70\pm5$ K, consistent with the magnetometric extrapolation of the low-field phase boundaries.

Field scans at 5, 40, 55, and 60 K are shown in Fig.~\ref{fig:sans}(d). The integrated SANS intensity first increases as the field selects a preferred sector or reorients helical domains, reaches a maximum at $H_{c1m}^{\mathrm{SANS}}$, and is then suppressed toward $H_{c2}^{\mathrm{SANS}}$. $H_{c1}^{\mathrm{SANS}}$ is treated as an onset interval of anisotropy rather than as a sharp intensity threshold. At $60\pm1$ K the perpendicular-to-field integrated intensity exhibits a maximum near $\mu_0H=0.025\pm0.005$ T. We use this field interval as a SANS-derived marker for the candidate A-phase region. This SANS signature occurs in the same field-temperature region as the magnetometrically defined susceptibility anomaly and is therefore treated as supporting evidence rather than as a standalone structural identification of an A phase.

\section{Combined SANS and magnetometry phase diagram}

\begin{figure}[htbp]
\centering
\includegraphics[width=\textwidth]{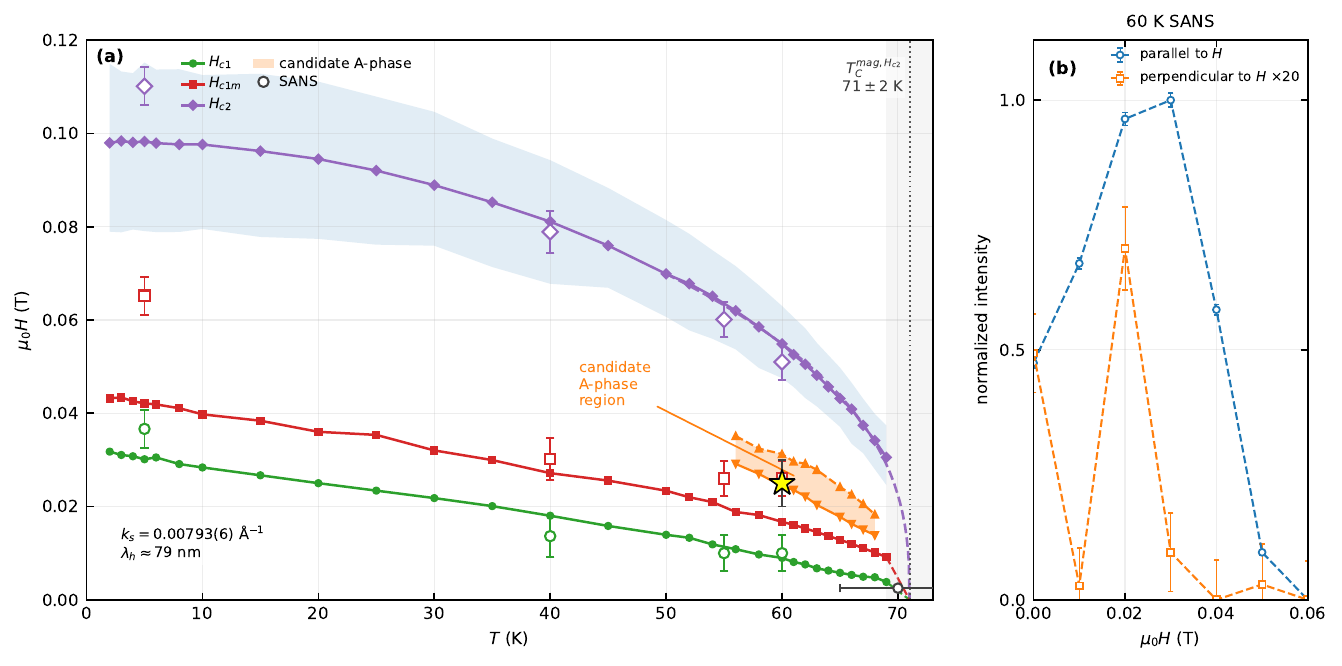}
\caption{(a) Combined SANS and magnetometry $H$-$T$ diagram. Magnetometric $H_{c1}$, $H_{c1m}$, and $H_{c2}$ are compared with SANS field estimates; the candidate A-phase region, the SANS-derived marker at 60 K and $\mu_0H=0.025\pm0.005$ T, and the $T_C^{\mathrm{SANS}}$ marker are included. (b) Normalized integrated SANS intensity at 60 K; the perpendicular-to-field component is multiplied by 20.}
\label{fig:combined}
\end{figure}

Figure~\ref{fig:combined}(a) combines the magnetometric phase boundaries, SANS critical fields, the candidate A-phase region, the SANS-derived marker for this region, and the zero-field SANS-derived magnetic ordering scale. Figure~\ref{fig:combined}(b) shows the field dependence of the integrated SANS intensity at 60 K, from which the SANS-derived marker at $\mu_0H=0.025\pm0.005$ T is obtained. The agreement is best understood at the level of the overall phase topology rather than by imposing point-by-point identity of each criterion. Magnetometry provides dense temperature coverage and sharply defined derivative criteria; SANS provides structural confirmation of the helimagnetic modulation and independent field scales at selected temperatures. Table~\ref{tab:fields} summarizes the quantitative comparison at the temperatures of the field-dependent SANS scans.

\begin{table}[htbp]
\caption{Comparison of characteristic fields obtained from magnetometry and SANS at the temperatures of the field-dependent SANS scans. All fields are $\mu_0H$ in Tesla. The SANS $H_{c1}$ value is given as an onset interval; the magnetometric values at 55 K are linearly interpolated between the 54 and 56 K isotherms.}
\label{tab:fields}
\begin{ruledtabular}
\begin{tabular}{ccccccc}
$T$ (K) & $H_{c1}^{\mathrm{mag}}$ & $H_{c1}^{\mathrm{SANS}}$ & $H_{c1m}^{\mathrm{mag}}$ & $H_{c1m}^{\mathrm{SANS}}$ & $H_{c2}^{\mathrm{mag}}$ & $H_{c2}^{\mathrm{SANS}}$ \\
5  & 0.030 & 0.033--0.041 & 0.042 & 0.065 & 0.098 & 0.110 \\
40 & 0.018 & 0.009--0.018 & 0.027 & 0.030 & 0.081 & 0.079 \\
55 & 0.011 & 0.006--0.014 & 0.020 & 0.026 & 0.064 & 0.060 \\
60 & 0.009 & 0.006--0.014 & 0.017 & 0.026 & 0.055 & 0.051 \\
\end{tabular}
\end{ruledtabular}
\end{table}

The combined diagram separates the relevant low-field temperature scales. The previously published bulk characterization of an independently synthesized sample gave $T_C=65.6$ K \cite{tsvyashchenko2026}. The present low-field magnetometric scale is $T_C^{\mathrm{mag},H_{c2}}=71$ K, while zero-field SANS gives $T_C^{\mathrm{SANS}}\approx70$ K. This modest offset is expected for samples obtained from independent high-pressure/high-temperature syntheses and should not be overinterpreted. The absolute magnetic parameters vary somewhat between syntheses, but the B20 helimagnetic phenomenology and the characteristic low-field scales remain reproducible.

\section{Discussion}

The combined results establish \FeRhSi{} as a 4d-substituted B20 helimagnet with a directly observed long-period modulation. The measured period $\lambda_h\approx79$ nm is substantially longer than in MnSi and in many \FeCoSi{} compositions, and is close to the long-period scale reported for FeGe \cite{muhlbauer2019,grigoriev2007brief,grigoriev2007field,wilhelm2011}. Rh substitution therefore appears to tune the balance of exchange and DMI in the B20 structure. Together with the previously reported spontaneous moment $m\approx0.48\,\mu_B$ \cite{tsvyashchenko2026}, the SANS result places \FeRhSi{} among itinerant chiral B20 helimagnets while adding a 4d-substituted member to this materials family.

The magnetometric phase diagram follows the expected B20 sequence of low-field helimagnetic-domain reorientation, a field-selected conical regime, and a crossover toward the field-polarized state at $H_{c2}$. The candidate A-phase region appears near the top of the helimagnetic dome and within $H_{c1}$-$H_{c2}$, similar to where A-phase phenomenology occurs in MnSi \cite{muhlbauer2009,reimann2018}, FeGe \cite{wilhelm2011}, \FeCoSi{} \cite{munzer2010}, and Cr$_{0.82}$Mn$_{0.18}$Ge \cite{ukleev2025}. The detailed comparison of $\chi(H)$ with field-dependent SANS is important here: the candidate region is associated with a depression of the differential susceptibility, a positive $d^2M/dH^2$ lobe, and a perpendicular-to-field SANS-intensity maximum in the same field-temperature range. We therefore shade only the restrictive positive-curvature region as a possible A-phase interval, without claiming that it uniquely defines the full stability range of such a phase. In a single crystal of \FeRhSi{} the boundaries, particularly on the low-field side, may vary with crystallographic orientation and demagnetizing conditions. Compared with the well-resolved A-phase or skyrmion-lattice regimes reported in MnSi \cite{muhlbauer2009,reimann2018}, \FeCoSi{} \cite{munzer2010,yu2010}, FeGe \cite{wilhelm2011}, and Cu$_2$OSeO$_3$ \cite{adams2012,seki2012}, the present evidence should be regarded as supporting evidence for a candidate A-phase region. Precursors, phase coexistence, and metastable states are known in B20 magnets \cite{bauer2016,reimann2018,bauer2012}, and similar effects may play a role here. SANS in the appropriate geometry \cite{muhlbauer2009,munzer2010}, complementary real-space probes \cite{yu2010}, or topological-Hall measurements \cite{schulz2012} will be needed to test for a skyrmion-lattice state.

The SANS result is central because it directly establishes the helical modulation that bulk magnetometry could only infer. The field scans further support the phase-boundary picture by showing the redistribution and eventual suppression of magnetic Bragg intensity \cite{muhlbauer2019,grigoriev2007field}. The different definitions of $H_{c1}$ are also physically sensible: magnetometry detects curvature changes in $M(H)$, whereas SANS detects the onset of anisotropic scattering \cite{muhlbauer2019,grigoriev2007field}. These observables track related but distinct aspects of the field-induced reorientation process; small offsets between the corresponding $H_{c1}$ estimates should therefore not be interpreted as contradictions.

Several aspects of the phase-boundary comparison should be interpreted cautiously. The sample is polycrystalline, the SANS data were taken in transverse geometry, and the SANS intensities are relative rather than absolutely calibrated. The $H_{c1}$, $H_{c1m}$, and $H_{c2}$ criteria therefore need not coincide point-by-point between magnetometry and SANS. More generally, disorder, spin fluctuations, and magnetic history are known to modify phase boundaries and metastability in B20 magnets \cite{munzer2010,bauer2016,ukleev2025,bauer2012}, and such effects may also influence \FeRhSi. Within these limitations, the agreement of the SANS-derived helimagnetic scale with $T_C^{\mathrm{mag},H_{c2}}$ and the placement of the perpendicular-to-field SANS-intensity maximum inside the magnetometric candidate region provide supporting evidence for the proposed low-field phase diagram.

\section{Conclusions}

We have combined low-field magnetometry and SANS to study the magnetic phase diagram of the B20 helimagnet \FeRhSi. SANS directly confirms long-period helimagnetic order with $k_s=0.00793$ \AA$^{-1}$, $\lambda_h\approx79$ nm, and $T_C^{\mathrm{SANS}}\approx70$ K. Magnetometry yields a low-field $H$-$T$ diagram with $H_{c1}$, $H_{c1m}$, and $H_{c2}$ and gives $T_C^{\mathrm{mag},H_{c2}}=71\pm2$ K as the temperature where the characteristic fields are expected to extrapolate to zero. Additional structure in $\chi(H)$, represented by a positive $d^2M/dH^2$ lobe inside $H_{c1}$-$H_{c2}$, identifies a continuous candidate A-phase region over approximately 56--68 K, while SANS shows a perpendicular-to-field SANS-intensity maximum at 60 K and $\mu_0H=0.025\pm0.005$ T in the magnetometrically defined region. Direct identification of a skyrmion lattice and its structural parameters in \FeRhSi{} remains an open task for future SANS geometries and complementary probes.

\begin{acknowledgments}
This research was funded by the Russian Science Foundation, Grant No. 25-12-68013 (22-12-00008-$\pi$).
\end{acknowledgments}

\bibliography{references}

\end{document}